\documentstyle[aps,pra,multicol,epsf]{revtex}
\begin{document}
\draft
\title
{Dissipation and Decoherence   in a Quantum Register
 }
\author {Paolo Zanardi\footnote{zanardi@isiosf.isi.it}}
\address{ Institute for Scientific Interchange  Foundation, \\Villa Gualino,
Viale Settimio Severo 65, I-10133 Torino, Italy\\
 and  Unit\`a INFM, Politecnico di Torino,\\
Corso Duca degli Abruzzi 24, I-10129 Torino, Italy
} 
\maketitle
 \begin{abstract}
{ A model for a quantum register $\cal R$ made of $N$  replicas
of a $d$-dimensional quantum system (cell) coupled with the environment,
is studied  by means of a Born-Markov
Master Equation (ME). Dissipation and decoherence are discussed in various
cases in which a sub-decoherent enconding can be rigorously found.
For the qubit case ($d=2$)  we have solved, for small $N,$ the ME by numerical direct integration
and studied, as a function of the coherence length $\xi_c$ of the bath,
fidelity
and decoherence rates of  states of the register.
For large enough $\xi_c$ the singlet states of the global $su(2)$ pseudo-spin algebra
of the register (noiseless at $\xi_c=\infty$) are shown to have a much smaller
deoherence rates than the rest of the Hilbert space.
} 
\pacs{71.10.Ad , 05.30.Fk}
\end{abstract}
\begin{multicols}{2}[]
\narrowtext
\section{Introduction}
Preserving  coherence in a quantum system is one of the most demanding features required
to be able to 
 take practical advantage from the implementation
of the    objects of  quantum information e quantum computation theory \cite{QC}.
Indeed all the additional power, with respect to the classical case, arising from the quantum nature
of the information-processing  device  depends on the
complex linear structure of the state space of a quantum system and on the invariance of such structure
under  (unitary) time-evolution.
The system is therefore endowed   with  
a massive intrinsic  parallelism and the capability of exhibiting
interference.
Unfortunately all of this holds only for {\sl closed} quantum systems.
Real systems are unavoidably coupled with  the environment in which they are enbedded,
hence they have to be considered as {\sl open} systems, no matter how weak is the interaction.
The relevant state manifold has now a {\sl convex structure} \cite{DAV},
the dynamics in  general  is no longer unitary and the interference patterns
may disappear. 
This set of effects is  known as the {\sl decoherence} problem \cite{ZUDI}.
The protection of quantum-encoded information against  environmental
noise has been, up-to-now,  mainly addressed in the framework  of the so-called
{\sl error correcting} codes \cite{ERROR}. These are essentialy schemes
to encode redundantly  information in  such a way that it can be recovered
also when (a few) 'errors' due to external sources have occured. 
Such schemes are often based   
on suitable measurement  protocols that have to be performed
frequently  enough to  keep the error level within the scope of the given encoding.
Of course  this implies that quantum information-processing systems
have to be coupled with a classical measurement apparatus:
even leaving aside the obvious pratical difficulties, such necessity naturally
 leads, at least, 
to a severe slow-down of the computational speed. 
More recently \cite{ZARA}( see also \cite{PALMA}\cite{LUGU})
has been put forward the idea that,  conceptually, a more efficient quantum state protection
can be realized by enconding the information in subspaces that the (non-unitary) dynamics
makes {\sl intrinsecally} more robust against the perturbation due to the environment. 
Here the attitude is, in some sense, opposite to that at the basis of  error correcting codes:
now one aims to encoding states that cannot be easily corrupted  rather 
than  to look for states that can be easily corrected. 
In this approach one has to assume explicit models of  system-environment interaction
and try to design the various ingredients  in  such a way that
that the algebraic-dynamical  structure of the global system gives rise to the stable subspaces one is looking for.
Since the typical environment consists of infinitely many  degrees of freedom
a direct Hamiltonian approach to the problem, is not the most suitable
except for some simplified situation \cite{ZANA}.
In this paper we address the problem of dynamically stable quantum enconding
within a  Master Equation formalism  that allows us 
to deal directly with the marginal dynamics of the computational degrees of freedom.
The relevant information about the environment is  contained in a few parameters 
appearing in the Master equation itself.
The system considered is the model of  a {\sl quantum register}:
$N$  replicas of a given finite-dimensional quantum system (the cell).
If the the cell is two-dimensional one obtains a $N$-{\sl qubit} register.
The key feature for the existence of the sub-decoherent codes
is the possibility of partioning the register in clusters (possibly
 coinciding with a single cell or  with the whole register) 
such that the cells within each cluster are collectively
perturbed by the environnent. It is the  dynamical symmetry of the cluster 
that allows to single out collective (entangled) states that, at least on a short time scale,
are unaffected by the noise and therefore evolve unitarily.
This mechanism  has a well-known counterpart in quantum optics
given by the phenomenon of {\sl subradiance} \cite{SUB}
\\
The paper is organized as follows:
in sect. II we introduce the model, in sect. III  are discussed the general features
of both  Master Equation and  the sub-decoherent codes.
The cases of purely dephasing and dissipative coupling 
with the environment  are analysed respectively in sect. IV and V.
Sect. VI contains some conclusions.
\section{ The Model}
We call a  system ${\cal R}$  a {\sl quantum register} with $N$ $d$-cells,
if ${\cal R}$ is  
composed by $N$ replicas of a $d$-level system.
The Hilbert space is given by ${\cal H}_{\cal R}=\otimes_{i=1}^N {\cal H}_i,$ where ${\cal H}_i\cong {\bf{C}}^d,
(i=1,\ldots,N)$
is the single $d$-cell  Hilbert space.
In particular, if $d=2$ one has a $N$-{\sl qubit} register.
The set of the {\sl states} (density matrices) of ${\cal R}$ is
\begin{equation}
{ S}_{\cal R}=\{ \rho\in\mbox{End}({\cal H}_{\cal R})\,\colon\,
\rho\ge 0,\; \rho=\rho^\dagger,\; \mbox{tr}^{\cal R}\,\rho=1\}.
\end{equation}
${ S}_{\cal R}$ is {\sl not} a linear subspace of $\mbox{End}({\cal H}_{\cal R})$ but a convex submanifold.
The register is coupled with an uncontrollable  environment  $\cal B$ (from now on the {\sl bath}).
The time-evolution of the states of the closed system ${\cal R} +{\cal B},$  is generated 
by a Hamiltonian of the form $H=H_{\cal R}+H_{\cal B}+H_{\cal I}.$
We now discuss the structure of each of these terms.
The bath will be considered as a single bosonic field, namely
$H_{\cal B}=\sum_k\omega_k\,b^\dagger_k\,b_k$ 
describes  a set of  non-interacting linear oscillator (field modes).
The self-hamiltonian ${\cal H}_{\cal R},$ of $\cal R,$
is assumed, for the time being, to be the sum of single-cell Hamiltonians $H^{C}_i$
(i.e. the register is an array of non-interacting cells).
The register-bath interaction is given by sum of  the bath-cell interactions
\begin{equation}
H_{\cal I}= \sum_{ki\alpha}( g_{ki\alpha} \,b^\dagger_k\,A^\alpha_i +\mbox{h. c}).\label{Hint}
\end{equation}
Here the    $A^\alpha_i$'s are single-cell operators
whose  action is non-trivial only on the $i$-th tensor factor of ${\cal H}_{\cal R}$,
representing the various interaction channels through which the $i$-th cell can be coupled with the bath.  
Although this kind of situation can be suitably handled by resorting to the notion of  dynamical algebra
\cite{SOBI},
in the following we shall assume that there is just one  dominant interaction, 
the corresponding operator being $A_i.$
As it is well-known the generic effect of $H_{\cal I}$ on the marginal dynamics 
of $\cal R$ is to induce  dissipation and decoherence.
The first effect, of course, consists in the irrevesible loss of register energy
into the bath. 
Decoherence is a pure quantum effect consisting in the destruction of phase coherence of the register states:
due to the entanglement with the bath the initial pure preparations of the register become 
in a very short time-scale mixed.
The interplay between these two phenomena is strictly related to the the nature of
the operators $\{A_i\}.$
Now we make another simplifying assumption, supposing that
 the $A_i$'s are 
eigenvectors
of  the adjoint action of $H_{\cal R},$
\begin{equation}
 [H_{\cal R},\,A_i]=-\epsilon \,A_i\quad(\epsilon\in{\bf{R}}_0^+).
\label{algebra}
\end{equation}
This  means that  
if $\epsilon>0,$ the necessarily non-hermitian and traceless, 
$A_i$'s ($A_i^\dagger$'s ) are the destruction (creation) operators of  {\sl elementary
cell excitations }of  $H_{\cal R}.$  
Notice that the energy $\epsilon$ does not depend on the cell index $i$,
in that we are considering replicas of the same physical system.
If one considers the zero temperature case, in which only the bath-vacuum is involved,
the  effect of $H_{\cal I},$ therefore will be that of letting the register relax to the
the $A$-vacuum $|A_0\rangle,$ ($A_i \,|A_0\rangle=0,\forall i$) by exciting the bath modes.
On the other hand, if $\epsilon=0$ the possibly hermitian  $A_i$'s belong to a symmetry algebra of $H_{\cal R},$
and no energy-exchange at all occurs: the effect of register-bath interaction
is pure decoherence.
 A quantity that will play
an essential role in 
the following  is  
the {\sl bath coherence length}
 $\xi_c,$
which, in an Hamiltonian approach, can be defined as the spatial
 scale over which the coupling constants $g_{ki}\equiv g_k(i),$ have a non-negligible
variation; when
$\xi_c=\infty,$ the $g_{ki}$'s no longer depend on the qubit index $i.$
This limit will be referred to as the {\sl replica symmetric} point, in that
for $\xi_c=\infty,$ the dynamics becomes invariant under the action of the symmetric group ${\cal S}_N$ of
the cell permutations and only the collective operators $A=\sum_i A_i$ are effectively
coupled with the bath.
This situation  corresponds to the well-known Dicke limit of quantum optics \cite{DICKE}
\\
To exemplify this situation let us consider the $\xi_c=\infty,$ limit with 
$A_i=\sigma_i^-.$ In this case, as far as the coupling with the environment is concerned,
  the relevant register operators are  
 $S^\alpha=\sum_{i=1}^N \sigma_i^{\alpha},\,(\alpha=\pm).$
Let  $H_{\cal R}= \epsilon\, S^z + H^1_{\cal R},$
where $H^1_{\cal R}$ is a qubit-qubit interaction term, and  suppose that the latter is
$su(2)$-invariant (i.e. $[H^1_{\cal R}, \,S^\alpha]=0,\,(\alpha=\pm,z)$);
one has then the commutation relation $[H_{\cal R},\,S^{\pm}]=\pm\epsilon\, S^{\pm}.$
There  follows that   for  large time the register relaxes to the lowest $S^z$-eigenstate allowed
by the total spin conservation. If  $ H^1_{\cal R}=0,$ this amounts to a ground-state
relaxation, whereas  if $\epsilon=0$ and $H^1_{\cal R}\neq 0,$ there is no energy loss. 
This example will be discussed with greater detail in sect. V.
\section{Master Equation}
The quantum dynamics of the system ${\cal R}+{\cal B}$ 
is highly non trivial, and exact results are difficult to obtain.
Nevertheless one is mostly interested in the register marginal dynamics
(i.e. forgetting about the bath degrees of freedom) in order to study  stability
against  external noise of the information-coding states of ${\cal R}$.
This issue can be conveniently addressed in the framework
of the Liouville-von Neumann equation for  open systems, the so-called Master equation (ME) .
Following the standard Born-Markov scheme where one   traces out the bath degrees of freedom,
which is assumed to be in the state $\rho_{\cal B},$ one obtains
 a closed  equation for the marginal density matrix of $\cal R,$ 
 of the form 
\begin{equation}
\dot\rho = {\bf{L}}(\rho)\equiv (i\,\mbox{ad}\,H_{\cal R}^\prime +\tilde{\bf{L}})(\rho),
\label{ME}
\end{equation}
 where as usual $\mbox{ad}\, H(\rho)\equiv [\rho,\,H],$ denotes the
adjoint action of $H.$
The  {\sl superoperator}  ${\bf{L}}$ is called the Liouvillian.
The action of the non-Hamiltonian (dissipative) part is 
\begin{eqnarray}
\tilde{\bf{L}}(\rho) &=& 
  \sum_{ij=0}^{N-1} \{ \Gamma_{ij}^{(-)}\,A_i\,\rho\,A_j^\dagger -\frac{\Gamma^{(-)}_{ji}}{2}\,( 
A_i^\dagger\,A_j\,\rho + \rho\,A_i^\dagger\,A_j ) \nonumber\\
 &+& \Gamma^{(+)}_{ij}A_i^\dagger\,\rho\,A_j -\frac{\Gamma^{(+)}_{ji}}{2}\,(
A_i\,A_j^\dagger\,\rho + \rho\,A_i\,A_j^\dagger)\},
\label{masterEq2}
\end{eqnarray}
where the $\Gamma_{ij}^{(\pm)}$'s are 
temperature dependent coupling constants
containing all relevant  information about the bath. 
They are respectively associated to the  process of de-excitation and excitation of the qubit system.
At $T=0,$ one has $\Gamma^{(+)}_{ij}=0.$
The renormalized Hamiltonian  $H_{\cal R}^\prime=H_{\cal R}+\delta H_{\cal R},$
where, by introducing the Lamb-shift parameters  $\Delta_{ij}^{(\pm)}$
\begin{eqnarray}
\delta H_{\cal R} =\sum_{ij} (\Delta_{ij}^{(-)} A^\dagger_i\,A_j+ \Delta_{ji}^{(+)} A_i\, A_j^\dagger).
\label{renorm}
\end{eqnarray}
At zero temperature the excitation terms $\Delta_{ij}^{(+)}$ are vanishing.
Notice that these terms make the cells interacting even though
$H_{\cal R},$ is a free-cell Hamiltonian. On the other hand, it follows 
from relation (\ref{algebra}),
 that $[\delta H_{\cal R},\,H_{\cal R}]=0;$ this means that, in this model, the Lamb shift
terms are not responsible for additional register energy loss, but they are a source 
of   dephasing.
Let $n_k =\mbox{tr}^{\cal B} (b^\dagger\,b_k\,\rho_{\cal B})$ be the mean occupation number of the mode
$k$ in the initial (thermal) bath state $\rho_{\cal B}$.
The explicit form for the coefficients appearing in the ME (\ref{ME}) is
\begin{eqnarray}
\Gamma^{(\pm)}_{ij}&=&\pi\,\sum_k g_{ki}\,\bar g_{kj}\, (\,n_k +\theta(\mp)\,)\,\delta(\omega_k-\epsilon),\nonumber \\
\Delta^{(\pm)}_{ij} &=&{\cal P}\sum_k\frac{g_{ki}\,\bar g_{kj}}{\omega_k-\epsilon}\,(\,n_k +\theta(\mp)\,).
\label{explicit}
\end{eqnarray}
$\theta$ is the customary Heaviside function, and ${\cal P}$ denotes the principal part. 
From these relations it follows that ${\bf{\Gamma}}^{(\pm)}$ and ${\bf{\Delta}}^{(\pm)}$
are hermitian. Furthermore ${\bf{\Gamma}}^{(\pm)}\ge 0$ and ${\bf{\Gamma}}^{(-)}\ge {\bf{\Gamma}}^{(+)}.$
It is important to notice that the assumption (\ref{algebra})
plays an essential role in the derivation of the ME, in that  it allows 
to move to  the interaction picture  (respect to $H_{\cal R}$)
$A_i\rightarrow A_i\, e^{-i\,\epsilon\,t}.$ 
This  is necessary in order to separate
the (fast) dynamics generated by the self-Hamiltonian from the (slow) one generated by the coupling
with the bath.
When only the collective cell-operators $A$ are coupled with the bath,
   $H_{\cal R}$ has to satisfy   condition (\ref{algebra}) 
only with respect to them. Given  such  $ H_{\cal R}$ one can obtain a family 
of new register Hamiltonians,  fulfilling the same constraint simply by adding 
terms commuting with $\{A,\,A^\dagger\}.$ 
Introducing the notation 
$A_i^\sigma =\theta(-\sigma)\,A_i +\theta(\sigma)\,A_i^\dagger,$
 equation (\ref{masterEq2}) can be cast in the compact form
\begin{equation}
\tilde{\bf{L}}(\rho)=\frac{1}{2}\sum_{ij,\sigma=\pm} \Gamma^{(\sigma)}_{ij}\, \left (
2\,\,A_i^\sigma \,\rho\,A_j^{-\sigma} -\{ A_j^{-\sigma}\,A_i^{\sigma},\,\rho\}\right ).
\end{equation}
Diagonalizing the hermitian matrices ${\bf{\Gamma}}^{(\sigma)}=\|\Gamma_{ij}^{(\sigma)}\|,\,(\sigma=\pm)$
one obtain the following canonical form for the dissipative part of the Liouvillian \cite{LIN}
\begin{equation}
\tilde{\bf{L}}(\rho)= \frac{1}{2}\sum_{\mu,\sigma=\pm} \lambda_\mu^\sigma\left (
[L_\mu^\sigma\,\rho,\, L_\mu^{-\sigma}]+ [L_\mu^\sigma,\,\rho\, L_\mu^{-\sigma}]\right ), 
\label{lindblad}
\end{equation}
where $\{\lambda_\mu^\sigma\}$ are the eigenvalues of $
{\bf{\Gamma}}^{(\sigma)}.$
Moreover  $L_\mu^\sigma=\sum_i u^\mu_i\,A_i^\sigma,$ $u_i^\mu$ denoting the components of the eigenvectors
of ${\bf{\Gamma}}^{(\sigma)}.$
The $L_\mu^\sigma$'s will be referred to as the Lindblad operators.
Given an initial pure preparation  $|\psi_0\rangle$ of the register, one  defines
$F(t)\equiv \langle\psi_0|\rho(t)|\psi_0\rangle$ {\sl fidelity}. 
Such quantity measures the degree of similarity with the initial preparation
that a state maintains during its time-evolution. 
 Another   quantity that one introduces in order to study the quantum coherence loss due to the bath is 
 $\delta(t)=\mbox{tr}\,(\rho(t)- \rho(t)^2),$ called {\sl linear entropy} (or idempotency deficit).
This quantity shares with the von Neumann entropy $S=-\mbox{tr} \rho\,\log \rho,$
the fundamental property $\delta[\rho]=0\Leftrightarrow \rho^2=\rho$ (i.e. they both vanish iff
$\rho$ is a pure state).
On the other hand, since the linear entropy does not involve trascendent operatorial functions, it
is much simpler to evaluate than $S.$
To characterize the degree of stability of the states
it is useful to consider the   short-times expansion
\begin{equation}
\delta(t)= \delta(0)+\sum_{n=1}^\infty \frac{1}{n!}\,(\frac{t}{\tau_n})^n.\label{expansion}
\end{equation}
In the following   $\tau_n$ ($\tau_n^{-1}$ ) will be referred  to as the $n$-th order decoherence time (rate). 
One  straightforwardly finds 
\begin{equation}
1/\tau_n^{n}=-\mbox{tr}^{\cal R}
\left \{
\sum_{k=0}^n \pmatrix{n\cr k}\,{\bf{L}}^{n-k}(\rho)\,{\bf{L}}^{k}(\rho) \right \}\quad (n\ge1).
\end{equation}
Since in the following $\tau_1$ will be play a major role, here we report explicitly the first  decoherence rate
\begin{equation}
\frac{1}{\tau_1}= -2\,\mbox{tr}\{ \rho\,\tilde {\bf{L}}(\rho)\}. 
\label{firstime}
\end{equation}
In particular, for  a pure initial preparation $\rho=|\psi\rangle\langle\psi|,$  one has $\delta(0)=0$ and 
\begin{equation}
\tau_1^{-1}[|\psi\rangle]=2\,\sum_{\mu,\sigma=\pm} \lambda_\mu^\sigma \left (
\langle\psi|\,L_\mu^{-\sigma}\,L_\mu^\sigma\,|\psi\rangle -|\langle\psi|\,L_\mu^\sigma\,|\psi\rangle|^2
\right ),
\label{firstime1}
\end{equation}
whereby one  notices that the Hamiltonian component of the Liouvillian does not contribute
to the first order decoherence time (this comes from 
$\mbox{tr}^{\cal R} \{ \rho\,\mbox{ad}\,H_{\cal R}^\prime(\rho)\}=0$).
Of course these expressions obtained within the ME equation formalism
(which relies on the Born-Markov assumption) differ from
the ones that one could get by the exact temporal evolution induced
by the interaction Hamiltonian (\ref{Hint}) (see for example \cite{KIM},
\cite{LUGU1}).
Nevertheless, as far as the issue of code stability classification   
is concerned,
this is not crucial in that the (exact) first order decoherence rate $1/\tau_1^{ex}$
is vanishing for pure initial state vanishing and $\tau_2^{ex}$
essentialy corresponds to $\tau_1.$
\subsection{codes}
The ME  with initial condition $\rho$
has formal  solution $\rho(t)=e^{t\,{\bf{L}}}(\rho),$ obtained by exponentiation of the Liouville super-operator
${\bf{L}}.$
The {\sl stationary} solutions $\rho(t)=\rho,$ are therefore the states 
belonging to $\mbox{ker}\,{\bf{L}},$
where $ \mbox{ker}\,{\bf{L}}=\{ \rho\in\mbox{End}({\cal H}_{\cal R})\,\colon\,{\bf{L}}(\rho)=0\}.$
When $\tilde{\bf{L}}(\rho)=0$ it follows, from equation (\ref{firstime1}),  that $\delta(t)=O(t^2)$
(whereas for the fidelity one finds $F(t)=1-O(t^2)$).
Such a state will be called {\sl sub-decoherent}.
In general the adjoint action of $H_{\cal R}$ maps sub-decoherent states
onto states such that $\tilde{\bf{L}}(\rho)\neq 0;$
 but when  ${ S}_{\cal R}\cap \mbox{ker}\,\tilde{\bf{L}}$
is $\mbox{ad}\, H_{\cal R}^\prime$-invariant  the Liouvillian evolution of each 
state $\rho\in\mbox{ker}\,\tilde{\bf{L}}$ becomes unitary: $\rho(t)=\exp (-i\,H_{\cal R}^\prime\,t)\,\rho\,
\exp (i\,H_{\cal R}^\prime\,t).$
 In particular one has $\delta(t)=0,\,\forall t>0$ (i.e. $\tau_n^{-1}=0,\,\forall n$).
This kind of state will be called {\sl noiseless.}
A subspace $\cal C\subset {\cal H}_{\cal R}$ such that each density matrix over it is a sub-decoherent (noiseless)
state will be referred to as a sub-decoherent (noiseless) {\sl code.}
\\
Let us suppose  $|\psi\rangle$ to be sub-decoherent.
First at all we notice that due to  non-negativity of matrices ${\bf{\Gamma}}^{(\sigma)}$
and from the Schwartz inequality,  each term of the sum in equation (\ref{firstime1})
is non-negative.
Therefore from $\tau^{-1}_1[|\psi\rangle]=0$  it follows that 
$\|L_\mu^\sigma\,|\psi\rangle\|^2= |\langle\psi|\,L_\mu^\sigma\,|\psi\rangle|^2,(\forall \mu,\sigma),$
which  in turn implies $|\psi\rangle$ to be a simultaneous eigenvector of {\sl all} the Lindblad operators.
Conversely if $|\psi\rangle$ is a simultaneous eigenvector of the $L_\mu^\sigma$'s then the
sub-decoherence constraint $\tau^{-1}_1=0$ is trivially fulfilled.
In other words  a necessary and  sufficient condition for the existence of  a sub-decoherent  code 
is the  existence of a simultaneous eigenspace of all  Lindblad operators $L_\mu^\sigma$
\begin{equation}
{\cal C}_\alpha=\{ |\psi\rangle\in{\cal H}_{\cal R}\,\colon\, L_\mu^\sigma\,
|\psi\rangle=\alpha^\sigma_\mu\,|\psi\rangle,\,
\forall \mu,\sigma\}.
\end{equation}
 The greater is $d[\alpha]\equiv\mbox{dim}\, {\cal C}_\alpha$ the more efficient is the encoding.
It is obvious that one has two quite different situations,  depending on whether 
or not  the $L_\mu^\sigma$'s are hermitian.
In fact, if $L_\mu^\sigma=(L_\mu^\sigma)^\dagger$ one has that the Lindblad operators commute,
$[L_\mu^\sigma ,\,L_\nu^\tau]=\sum_{ij} u_i^\sigma\, u_j^\tau \,[A_i,\, A_j]=0 ;$ then there exists a non trivial ${\cal C}_\alpha,$
furthermore ${\cal H}_{\cal R}=\oplus_\alpha {\cal C}_\alpha.$
On the other hand, if $L_\mu^\sigma\neq(L_\mu^\sigma)^\dagger$ the Lindblad operators
no longer span an abelian algebra and cannot to be simultaneously diagonalized.
 The only candidate as sub-decoherent code is
${\cal C}=\cap_{\mu\sigma} \mbox{ker}\,L_\mu^\sigma.$
Indeed the Lindblad operators satisfy  relation (\ref{algebra}),
from which one derives that the only allowed eigenvalue is $\lambda_\mu^\sigma=0.$
The proof is as follows \cite{NOTE}:
let $\{ |E_i\rangle\}_{i=1}^D $ a $H_{\cal R}$-eigenstates basis of ${\cal H}_{\cal R}$ ($H_{\cal R}\,|E_i\rangle= E_i\,|E_i\rangle,\,
E_{i+1}\ge E_{i},\,D =d^N$).
Since the $L_\mu^+$ are raising operators over the spectrum of $H_{\cal R}$ 
one has $L_\mu^+\,|E_i\rangle\propto |E_{i^\prime}\rangle,$ where $ i^\prime > i,\,E_{i^\prime}=E_{i}+\epsilon$ 
(in particular the maximum eigenvalue vector $|E_D\rangle$ is annihilated by $L_\mu^+,\,(\forall \mu)$).
Let $|\psi\rangle=\sum_{i=1}^D c_i\,|E_i\rangle$ an eigenvector of $L_\mu^+$
with eigenvalue $\lambda\neq 0$;
then one must have
 $L_\mu^+\,|\psi\rangle =\sum_{i=1}^{D-1}c_i L_\mu^+\,|E_i\rangle=\lambda \sum_{i=1}^{D} c_i\,|E_i\rangle$
hence $c_1=0.$  Acting on $|\psi\rangle$ with increasing powers of $L_\mu^+$ one analougously finds
 $c_{2}=c_{3}=\ldots=c_{D}=0,$ therefore if $\lambda\neq 0$ one would have $|\psi\rangle=0.$  \\ 
Let $\cal L$ the Lie algebra generated by the $L_\mu^\sigma$'s (i.e. the minimal subspace
of operators closed under commutation containing $\{ L_\mu^\sigma\}_{\mu\sigma}$)
then the code $\cal C$ is nothing but the {\sl singlet sector} of $\cal L;$
each $|\psi\rangle\in{\cal C}$ is a one-dimensional representation  space of $\cal L.$ 
From the   general form of the Lindablad operators one has   ${\cal L}\subset  \oplus_{i=1}^N {\cal L}_i$
where ${\cal L}_i$ is the (local) Lie algebra generated by the $A_i^\sigma$'s.
Generically one has ${\cal L}_i\cong \mbox{sl}\,(d,\,{\bf{C}}),$ therefore if the above inclusion 
is not strict it follows that
${\cal C}=\{0\},$ which has no use for quantum encoding. 
In order to obtain meaningful codes one has to impose  constraints 
on the algebraic structure generated by the Lindblad operators. 
The smaller $\cal L$ the easier will be the task of finding (by representation theory) non trivial $\cal C.$ 
Notice that, given such a sub-decoherent code, if $H_{\cal R}^\prime$ belongs
to the universal enveloping algebra ${\cal U}({\cal L})$ then  $\cal C$ is also necessarily noiseless.
\\
The matrices ${\bf{\Gamma}}_{ij}^{(\sigma)}$ and ${\bf{\Delta}}_{ij}^{(\sigma)}$  encode all the information
about the spatial correlations among the register cells induced by coupling with the  bath.
The actual form of these correlations depends (see equation(\ref{explicit}))
on the detailed form of the coupling functions $g_{ki},$
on the bath density of the state and  on temperature as well.
\\
Leaving aside strongly model-dependent
considerations and in view of keeping the  form of the ME here considered
as general as possible, in the following 
the matrices ${\bf{\Gamma}}^{(\pm)},\,{\bf{\Delta}}^{(\pm)},$
will be considered rather  as {\sl a priori} data
of the problem defining the basic dynamical equation (\ref{ME}).
In other words they are treated  as parameters that have to be 'engineered' in order
to  realize an advantageous  situation for quantum-enconding.
In this context the bath coherence length $\xi_c$ is better defined 
in relation to the spatial behaviour of the $\Gamma_{ij}^{(\pm)}.$
One can consider the following particular regimes, corresponding
to different 'effective' topologies of $\cal R.$
It is just from these topologies that constraints on the algebraic structure arise.
\begin{itemize}
\item[i)] ${\bf{\Gamma}}_{ij}= \Gamma\,\delta_{ij},\,(\forall i,j):$
this is the cell limit;
 the decoherence process occurs independently
in each cell.
The Lindblad operators coincide with the  $A_i^\sigma$'s
(${\cal L}=\oplus_{i=1}^N{\cal L}_i$).
\item[ii)] ${\bf{\Gamma}}_{ij}= \Gamma\,(\forall i,j):$
this is the replica symmetric point; the decoherence is collective.
The matrices ${\bf{\Gamma}}^{(\sigma)}$ have constant entries,
the only non-zero eigenvalue is $N$ and the corresponding Lindblad operators
are given by $L^\pm= N^{-1/2} \sum_i A^\pm_i$ (${\cal L}\cong {\cal L}_i$).
\end{itemize}
The limit i) is the one usually considered in Error Correction literature.
The case ii) corresponds to the so called Dicke limit of quantum optics

An interesting intermediate case between i) and ii) is
when the register is partioned in clusters such that in each cluster the cells are coupled
in the same way with the environment and different clusters are far enough to feel
scorrelated  environments. In other terms,
 if $l$ ($L$) is the typical intra-cluster (inter-cluster) distance,
we are supposing $l\ll\xi_c$ ($L\gg \xi_c$).
More formally we assume that there exists a partition  $\{C_\lambda\}_{\lambda=1}^M$
of the cell index set ${\bf{N}}_N,$ such that
\begin{itemize}
\item[iii)]
$\Gamma_{ij}=\Gamma_0$ if $ i,j\in C_\lambda,$
$0$ otherwise.
The Lindblad operators are the cluster-ones 
$L_\lambda^\sigma= N_\lambda^{-1/2}\sum_{i\in C_\lambda} A_i^\sigma,$
being $N_\lambda$ the number cells in the $\lambda$-th cluster (${\cal L}\cong \oplus_{i=1}^M {\cal L}_i$).
When $M=N$ and $M=1$ we recover respectively the cases i) and ii).\end{itemize}
For a clustered register the dynamics is invariant under the  action of the 
group  ${\cal G}\equiv {\cal S}_{m_1}\times\cdots\times {\cal S}_{m_M}\subset {\cal S}_N;$
at the replica symmetric point (cell limit) one has ${\cal G}={\cal S}_N$ (${\cal G}=\{{\bf{1}}\}$).
Some comments are now in order.
When the relation (\ref{algebra}) holds we see that both in the hermitian and in the non-hermitian
case the self Hamiltonian leaves invariant the code; nevertheless also in this rather special situation,
due to the renormalizing terms  (\ref{renorm}) sub-decoherence does not necessarily imply
noiselessnes. The point is that the ${\bf{\Gamma}}^{(\pm)}$'s  and the ${\bf{\Delta}}^{(\pm)}$'s
in general cannot be diagonalized simultaneously.
This can be understood, for example, by looking at the explicit form (\ref{explicit}):
 in the matrix elements ${\Delta}^{(\pm)}_{ij}$ appears a sum over {\sl all}  the bath modes 
whereas in the $\Gamma_{ij}^{(\pm)}$'s  are involved only the modes degenerate with 
the single cell eigenvalue $\epsilon.$
On the other hand, we see, from equation (\ref{explicit}), that the leading contribution to $\Delta_{ij}^{(\pm)}$
comes from  the same bath modes involved in $\Gamma_{ij}^{(\pm)},$ therefore assuming that
$\Delta_{ij}^{(\pm)}$ and $\Gamma_{ij}^{(\pm)}$ have the same structure can be in many cases a good approximation.
When this is the case also   $\delta H_{\cal R}$ can be written in terms of the Lindblad operators, namely
each sub-decoherent code ${\cal C}_{\alpha}$ is necessarily noiseless.
\section{ Decoherent coupling}
In this section we consider,
the   case in which the single cell-operators $A_i$ in equation (\ref{masterEq2})
are hermitian. 
Although this case is essentialy well-known we think that it is worthwhile to 
analyse it in that its exact solvability allows us  to shed some light onto the general features of the decoherence
process of many replicas of a given system coupled with the same environment. 
Here the ME is considered the starting point of the analysis,
we do not assume any {\sl a priori}  relation like equation (\ref{algebra}).
For the time being we set $H_{\cal R}=0.$
Let $|\alpha\rangle\equiv|\alpha_1,\ldots,\alpha_N\rangle$ denote a simultaneous 
eigenvector of the $A_i$'s with  $A_i\,|\alpha\rangle=\alpha_i\,|\alpha\rangle, (i=1,\dots,N).$
The operators $|\alpha\rangle\langle\alpha^\prime|$ are eigenvectors of the  Liouvillian
\begin{eqnarray}
{\bf{L}}(|\alpha\rangle\langle\alpha^\prime|)& =& W(\alpha,\alpha^\prime)\,|\alpha\rangle\langle\alpha^\prime|,
\nonumber \\
W(\alpha,\alpha^\prime) &=& i\,( \|\alpha\|^2_{\bf{\Delta}} -\|\alpha^\prime\|^2_{\bf{\Delta}}) 
-\|\alpha-\alpha^\prime\|^2_{\bf{\Gamma}},\label{eigeOp}
\end{eqnarray}
where $\|\beta\|^2_{\bf{M}}=\langle \beta, {\bf{M}}\,\beta\rangle,\,({\bf{M}}={\bf{\Delta}},{\bf{\Gamma}}\equiv\sum_{\sigma=\pm}{\bf{\Gamma}}^{(\sigma)}
\in\mbox{End}({\bf{C}}^N),\beta\in{\bf{C}}^N).$
Notice that $\|\bullet\|_{{\bf{M}}},$ is a semi-norm only if ${\bf{M}}\ge 0,$
and a norm only if $\mbox{ker}\,{\bf{\Gamma}}=\{0\}.$
Each state over ${\cal H}_{\cal R}$ can be written in the form
$\rho=\sum_{\alpha,\alpha^\prime} R_{\alpha,\alpha^\prime}\,|\alpha\rangle\langle\alpha^\prime|,$ 
therefore the general solution of equation (\ref{masterEq2})  is 
\begin{equation}
\rho(t)= \sum_{\alpha,\alpha^\prime} R_{\alpha,\alpha^\prime}\, e^{W(\alpha,\alpha^\prime)\,t}
 \,|\alpha\rangle\langle\alpha^\prime|,
\end{equation}
whereby one derives the following expressions for fidelity and linear entropy
\begin{eqnarray}
F(t)&=& \sum_{\alpha\alpha^\prime}|R_{\alpha\alpha^\prime}|^2 e^{W(\alpha,\alpha^\prime)\,t}, 
\nonumber\\
\delta(t)&=&1-\sum_{\alpha\alpha^\prime}|R_{\alpha\alpha^\prime}|^2 e^{2\,\Re\,W(\alpha,\alpha^\prime)\,t}.
\label{F-d}
\end{eqnarray}
By 
 equation (\ref{eigeOp})
the set of sub-decoherent and noiseless solutions of  Liouville equation (\ref{masterEq2})
is obviously
 related to the properties
of the matrices ${\bf{\Delta}}$ and ${\bf{\Gamma}}.$
First at all notice that from the second of equations (\ref{F-d}),  the immaginary terms in equation (\ref{eigeOp})
play no role in decoherence (in the restricted  meaning): indeed  they give rise
to the unitary transformation
 \begin{equation}
U_{ {\bf{\Delta}} }(t)=e^{-i\,t\,\delta H_{\cal R}}=
\sum_{\alpha} e^{i\, \|\alpha\|^2_{\bf{\Delta}} \,t}\,|\alpha\rangle\langle\alpha|.
\label{Lamb}
\end{equation}
It is straightforward to verify that the linear entropy is a monotonic non-decreasing function 
of  time, indeed
\begin{equation}
\dot\delta(t)=2\,\sum_{\alpha\alpha^\prime}|R_{\alpha\alpha^\prime}|^2  \,
\|\alpha-\alpha^\prime\|^2_{ {\bf{\Gamma}} }\ge 0,
\end{equation}
the  inequality following from the non-negativity of ${\bf{\Gamma}}.$
Cases i) and ii) imply, from $W(\alpha,\alpha)=0,$ that the diagonal states 
$\rho_\alpha\equiv |\alpha\rangle\langle \alpha|$ are fixed points of the Liouvillian evolution.
Furthermore
if $\alpha-\alpha^\prime\in\mbox{ker}\, {\bf{\Gamma}},
$ one has that the real part of $W(\alpha,\alpha^\prime)$  vanishes.
Case i) corresponds  to a solution that one could obtain assuming that each cell
is interacting with its own independent environment.
From equation (\ref{eigeOp}) it follows that the maximum decay rate is $O(N).$
In case i) $\mbox{ker}\, {\bf{\Gamma}}=\{0\}$ and only $\alpha=\alpha^\prime$
survives.
If the single-cell eigenvalues $\alpha_i$ are non-degenerate the eigenspace ${\cal H}(\{\alpha_i\})$
is one-dimensional and therefore  useless for quantum encoding.
If instead the $\alpha_i$'s are $m_i$-fold degenerate, then 
$d[\alpha]\equiv\mbox{dim} {\cal H}(\{\alpha_i\})=\prod_{i=1}^N m_i.$
The  density matrix corresponding to  $|\psi\rangle\in{\cal H}(\{\alpha_i\})$
evolves according the unitary transformation $U_{ {\bf{\Delta}} }(t)$:  
 these states are  noiseless.
The largest dimension for the {\sl noiseless} code ${\cal H}(\{\alpha_i\})$
is obtained for $\alpha_i=\alpha_M,\,(\forall i)$ where $\alpha_M$
is the single-cell eigenvalue with the maximum degeneracy.
In the qubit case $A_i=\sigma_i^z$ and $\alpha_i=\pm 1/2.$
In case even ${\bf{\Delta}}^{(\pm)}$ is proportional to the unit matrix,
then the unitary transformation (\ref{Lamb}) becomes trivial, 
being $\sum_i\alpha_i^2=N/4,\,(\forall\alpha).$
For the initial state $|\psi_0\rangle=2^{-N/2}\sum_\sigma|\sigma\rangle,$
uniform linear superposition of all the basis states, one can obtain explicit
analytical expressions for the linear entropy and the fidelity
\begin{eqnarray}
\delta(t)&=&1-e^{-\Gamma\,N\,t}\,\cosh^N(\Gamma\,t),\nonumber \\
F(t)&=&e^{-\Gamma/2\,N\,t}\,\cosh^N(\Gamma/2\,t).
\end{eqnarray}  
For $t\rightarrow\infty$ one finds $F\sim 2^{-N}$ and $\delta\sim 1-2^{-N},$
results that can be immediately understood from
$\rho(\infty)=2^{-N}\, \sum_\sigma |\sigma\rangle\langle \sigma|.$
Let us  turn to the case ii).
The operator $A= \sum_i A_i$ plays the role of {\sl pointer observable}\cite{ZURE}:
the diagonal elements with respect to its eigenstates basis of the density matrix
do not decohere, whereas the off-diagonal decay with a rate that is proportional
to their distances from the diagonal.
Now $\mbox{dim}\,\mbox{ker}\, {\bf{\Gamma}}=N-1,$ and the no-damping 
condition becomes $\sum_i\alpha_i=\sum_i\alpha_i^\prime.$
This means that in that case  the space
${\cal H}_\alpha$  spanned by the set $B_\alpha= \{ |\alpha\rangle\,\colon\, \sum_{i=1}^N\alpha_i=\alpha\}$
is decoherence-free.
In passing we note that, since $A$ is an extensive observable, at the replica-symmetric point the maximum decay rate
is $O(N^2).$
\\
In case iii)   the matrix ${\bf{\Gamma}}$ is block constant
and $\alpha-\alpha^\prime\in\mbox{ker}{\bf{\Gamma}},$ iff
$\sum_{j\in C_\lambda} \alpha=\sum_{j\in C_\lambda} \alpha_j^\prime\,(\lambda=1,\ldots,M).$ 
Now the relevant operators are the cluster operators
$L_\lambda=\sum_{j\in C_\lambda}A_i,$
the states built over a simultaneous eigenspace of the $L_\lambda$'s evolve in a noiseless way.
As usual, the situation is best exemplified by the qubit case.
Let us assume  that $A_i=\sigma^z_i,$ and $N$ even. 
At the $\xi_c=\infty$ point
the  most efficient noiseless-enconding is obtained  building states
over the eigenspace $S^z=0$. 
If ${\bf{\Gamma}}$ is partioned in blocks of $m$ (even) elements
one can encode in the subspace with zero cluster $z$-spin. Such code
has dimension
\begin{equation}
d(M)= \pmatrix{ m\cr m/2}^{M}.
\end{equation}
This enconding, with $m=2$ is essentialy that proposed in \cite{LUGU1}.
Till now we have supposed that the self-Hamiltonian were vanishing.
 If this is not the case,
one has that for an initial noiseless preparation
the state evolves infinitesimally in a unitary fashion.
For finite time the (possible) non-commutativity between $H_{\cal R}$ and the relevant Lindblad
(cell, cluster, register)  operators, destroys the coherence of $\rho.$
When relation (\ref{algebra}) holds ($\epsilon=0$) 
 $H_{\cal R}$ commutes with the  Lindbald operators. Working in a basis that simultaneously 
 diagonalizes  $H_{\cal R}$ and the $A_i^\sigma$'s one sees  
that  $U_{ {\bf{\Delta}} }(t)\rightarrow\exp (-i\,t\,H_{\cal R}),$ 
the evolution will remain unitary for finite times;  the initial pure states
never get mixed.
\section{Dissipative Coupling}
In this section we consider the case of non-hermitian $A_i;$ namely the case when the relation
(\ref{algebra}) holds with $\epsilon> 0.$
At zero temperature the eigenvalues $\lambda^+_\mu$ are vanishing. 
On the other hand  since 
 ${\bf{\Gamma}}^{(-)}\ge 0,$  $\lambda_\mu^-\ge 0,\, (\forall \mu)$  
one can immediately check that the register energy $E_{\cal R}(t) =\mbox{tr}^{\cal R}\left (
\rho(t)\,H_{\cal R}\right )$ is a monotonic non-increasing function.
Indeed
\begin{eqnarray}
\dot E_{\cal R}(t) &=& \mbox{tr}^{\cal R}\left (\tilde{\bf{L}}(\rho) \,H_{\cal R}\right) \nonumber \\
&=& -\epsilon\,\sum_\mu \lambda^-_\mu \,\mbox{tr}^{\cal R}\left ( L^+_\mu\,L_\mu^-\,\rho \right )\le 0,
\end{eqnarray}
where we have used
the irrelevance of Hamiltonian component of ${\bf{L}}$ (that is 
 $\mbox{tr}^{\cal R}\left (H_{\cal R},\,[H^\prime_{\cal R} ,\,\rho]\right)=0,$)
the relation $[L_\mu^\sigma\,L_\mu^{-\sigma},\,H_{\cal R}]=0,$ (that follows
form equation (\ref{algebra}) which holds for the Lindblad operators as well),
 and the non-negativity
of  operators $ L^+_\mu\,L_\mu^-$ and $\rho.$
As it has been observed in sect. III in the present case a sub-decoherent
code can be obtained if ${\cal C}\equiv\cap_\mu \mbox{ker}L_\mu\neq\{0\}.$
\\
Restated in this  formalism the essence of the result of reference  \cite{ZARA}
for the qubit case
is that at the $\xi_c=\infty$ point the Lindblad operators (and the renormalized 
self-Hamiltonian) belong
to a $N$-fold tensor representation of a semisimple (dynamical) Lie algebra,
out of which a non-trivial $\cal C$ can be built 
when $N$ is large enough.
In the  cell limit i)  if one can find a subspace ${\cal C}_i\subset {\cal H}$
annihilated by both $A^{(+)}_i$ and $A^{(-)}_i$  then 
${\cal C}\equiv {\cal C}_1\otimes\cdots\otimes  {\cal C}_N.$
An analog construction can be made in the cluster limit.
An important example is given by the qubit case.
One can design a register that supports noiseless encondings
if one is able to build $\cal R$ in such a way that iii) is satisfied with
$m=4$ qubits for cluster.
Then, according reference to \cite{ZARA}, a logical qubit can be encoded
 in each cluster.
It is  important to note that 
 the dimension of ${\cal C}$ decreases passing from ii) to iii), and from
iii) to i).
\\
In general one has $\Gamma_{ij}^{(\pm)}=\Gamma^{(\pm)}(i,\,j).$
The first order time-scale $\tau_1$ is a functional of $|\psi\rangle,$ depending on $\xi.$
The  optimal  states, with respect to 
the storage reliability on short times, are those  that minimize this functional
for a given bath coherence length.
Let us assume that $\Gamma_{ij}=\Gamma_0\,\gamma_{\xi} (i-j),$ where $\gamma_\xi(x)
\rightarrow 1,$ when $\xi_c\rightarrow\infty$ and $\gamma_\xi(x)
\rightarrow \delta_{x,0},$ when $\xi_c\rightarrow 0^+.$
The latter situation corresponds to the case in which each cell is coupled with
an independent bath, therefore 
 $\xi_c\in(0,\infty)$ interpolates between the independent bath
limit i) and the infinite coherence length bath case ii).
\subsection{qubit case}
Now we specialize to the $d=2$ case: $A_i^\pm=\sigma_i^\pm.$ 
Let the self-Hamiltonian be of the form 
$
H_{\cal R}=\epsilon\,S^z+H^1_{\cal R},
$
where the second term is a qubit-qubit interaction. In quantum computation
applications such a kind of term  might arise, for example,
 during the gate  processing.
If we assume that $[H^1_{\cal R},\,S^\alpha]=0,\,(\alpha=z,\pm)$ 
then equation (\ref{algebra}) holds.
Now we briefly recall the result of reference \cite{ZARA} 
at the replica symmetric point.
When $\xi_c=\infty,$ one finds that:
\begin{itemize}
\item[ i)] the total spin operator $S^2=(S^z)^2+1/2\{S^+,\,S^-\},$ is a constant of the motion,
\item[ii)] defining in the obvious way a ${\cal S}_N$-action $T$  over the density matrices manifold 
$S_{\cal R},$
one has $T_\sigma\,{\bf{L}}\,T_\sigma^\dagger={\bf{L}},(\forall \sigma\in {\cal S}_N),$ 
\item[iii)] the Lie algebra $\cal L$ generated by
the Lindblad operators $S^{\pm}$ is nothing but the global $su(2)$.
\item[iv)] Since at $\xi_c=\infty$ the coupling functions $g_{ki}$ are assumed to be {\sl strictly} qubit independent
also the Lamb-shift matrices $\Delta_{ij}^{(\pm)}$ have constant entries (i.e. $
\Delta_{ij}^{(\pm)}=\Delta_0^{(\pm)} \forall i,j$).
The renormalizing term can then to be written as 
$\delta H_{\cal R}= \Delta_0^-\, S^+\,S^- + \Delta_0^+\, S^-\,S^+$ 
\end{itemize}
From i)--iv) it follows that the Hilbert space ${\cal H}_{\cal R}$ splits dynamically according
the Clebsch-Gordan decomposition
of the $n$-fold tensor representation of $su(2)$
\begin{equation}
{\cal H}_{\cal R}= \oplus_{S=S_{min}}^{N/2}\oplus_{r=1}^{n_N(S)}{\cal H}_r(S),
\end{equation}
where $S_{min}=0$ ( $S_{min}=1/2$) if $N$ is even (odd).
The  subspace ${\cal H}_r(S)$ 
is an irreducible $su(2)$-module corresponding to the total spin eigenvalue $S\,(S+1),$
the latter occurring with multiplicity
\begin{equation}
 n_N(S)=\frac{(2\,S+1)\,N!}{(N/2+S+1)!\,(N/2-S)!}.
\label{mult}
\end{equation}
The general state over ${\cal H}_r(S)$ has the form:
$\rho=\sum_{M,M^\prime=-S}^S\rho_{M,M^\prime}\,|S M\rangle\langle S M^\prime|,$
where $S^2\,|S M\rangle=S\,(S+1)\,|S M\rangle,\,S^z\,|S M\rangle= M\,|S M\rangle,\,(M=-S,\ldots,S)$ and 
analogously for $|S M^\prime\rangle$.
For a pure state one has
\begin{eqnarray}
\tau_1(\infty)^{-1}&=& 2\,{\Gamma^{(-)}_{0}}\left (\langle\psi|\,S^+\,S^-\,|\psi\rangle
-|\langle\psi|\,S^-\,|\psi\rangle|^2\right )\\
&+ &2\,{\Gamma^{(+)}_{0}}\left (\langle\psi|\,S^-\,S^+\,|\psi\rangle
-|\langle\psi|\,S^+\,|\psi\rangle|^2\right ).
\end{eqnarray}
In particular, if $|\psi\rangle=|S M\rangle$ one obtains $(2\,\tau_1)^{-1}=\Gamma_0\,C^2_-(S,\,M)+
\Gamma^{(+)}_0\,C^2_+(S,\,M),$
where $C^2_\pm(S,\,M)=S\,(S+1)-M\,(M\pm 1).$
Let us consider the zero-temperature case ($\Gamma^{(+)}_{0}=0$) when 
only the de-excitation processes with strength
proportional to $\Gamma^{(-)}_{0}$ are active.
If $|\psi\rangle$ is a {\sl lowest-weight} spin state (i.e. $S^-\,|\psi\rangle=0$), 
one has $\tau_1(\infty)=\infty.$
This result is true for all the decoherence-times $\tau_n.$
At finite temperature the (excitations) terms weighted by $\Gamma^{(+)}_0$ are present as well.
On the $su(2)$-{\sl singlets} 
$|\psi\rangle\in {\cal C}\equiv  \oplus_{r=1}^{n_N(0)} {\cal H}_r(0),$ one has 
$S^+\,|\psi\rangle=S^-|\psi\rangle=0,$ and $\delta H_{\cal R}\,|\psi\rangle=0;$
 furthermore from the $su(2)$-invariance 
of $H^1_{\cal R}$ there follows that the unitary part  of  $\bf L$ maps the singlet sector  onto itself,
namely $\cal C$ is noiseless.
From equation (\ref{mult}) it follows that the minimum cluster  size to encode
a noiseless logical qubit is $N=4.$
Defining (in  obvious binary notation) the states
$
|A\rangle \equiv |0011\rangle +|1100\rangle,\; |B\rangle \equiv|0110\rangle+|1001\rangle,\;
|C\rangle \equiv|1010\rangle +|0101\rangle,
$
an orthonormal basis of ${\cal C}$ is given by
\begin{eqnarray}
|{\bf{0}}\rangle &\equiv & 2^{-1}\,(  |B\rangle -|A\rangle ), \nonumber \\
|{\bf{1}}\rangle &\equiv& 3^{-1/2}\,(|C\rangle -{2}^{-1}\,|A\rangle -{2}^{-1}\,|B\rangle).
\end{eqnarray}
If $H_{\cal R}^1=0$ this two states are energy degenerate, for non-vanishing
qubit-qubit interaction the degeneracy is lifted.
For example if 
\begin{equation}
H^1_{\cal R}= J\sum_{<ij>} \{ \sigma^z_i \sigma^z_j +1/2\,(\sigma_i^+\,\sigma_j^-+\sigma_i^-\,\sigma_j^+)\},
\end{equation}
is a Heisenberg  coupling between nearest neighbour qubits arranged on a
ring topology, one finds that $|{\bf{0}}\rangle$ and $|{\bf{1}}\rangle$ are
energy eigenstates with eigenvalues respectively given by $E_0=J$ and $E_1=-J.$
Since  $H^1_{\cal R}$ is  $su(2)$-invariant it is always possible to choose the singlet 
$|\psi\rangle$ among its eigenvectors.
It should be emphasized that the $su(2)$ singlet sector
is  noiseless for a wider class of ME's, with 
 Lindblad operators (and  self-Hamiltonian) given by arbitrary functions
of the global operators $S^\alpha ,\,(\alpha=\pm,z) $\cite{SYM}.
Indeed if these operators have the form
\begin{equation}
X=c_1\,{\bf{I}}+ F(\{S^\alpha\})
\label{UEA}
\end{equation}
where $F$ is an arbitrary operator-valued analytic function,
then -- since $F\,|\psi\rangle=0, (\forall \,|\psi\rangle\in
{\cal C}_N)$-- one obtains $X|_{{\cal C}_N}=c_1\,{\bf{I}}.$
This latter condition is sufficient to preserve the sub-decoherence
of ${\cal C}_N.$
Another way to understand  this result is that the operators
described by Eq. (\ref{UEA}) coincide with the ${\cal S}_N$-invariant sector
(symmetic subspace) of $\mbox{End}({\cal H}_{{\cal R}})$ .
Since ${\cal C}_N$ is an irreducible ${\cal S}_N$-module
from the Schur lemma it follows that $X|_{{\cal C}_N}\propto {\bf{I}}.$
\\
Turning back to the general case $\xi\in(0,\,\infty),$
if $N=2,$ for the initial states $ |\uparrow\uparrow\rangle,\,
|\downarrow\downarrow\rangle,\,
|\psi_{t,s}\rangle = 2^{-1/2}\,(|\uparrow\downarrow\rangle\pm|\downarrow\uparrow\rangle), 
$ one immediately finds
\begin{eqnarray}
\tau_1^{\uparrow\uparrow}(\xi) &=&(2\,\Gamma_0^{(-)})^{-1},\quad 
\tau_1^{\downarrow\downarrow}(\xi)=(2\,\Gamma_0^{(+)})^{-1},\nonumber\\
\tau_1^{t,s}(\xi)&=&( 2\,(\Gamma_0^{(-)}+\Gamma_0^{(+)})(1\pm \gamma_\xi(1)))^{-1}.
\end{eqnarray}
These equations show that in the generic case ($\Gamma^{(\pm})\neq 0$)
for finite coherence length $\xi$ all the first order decoherence times  are finite as well,
whereas for $\xi\rightarrow \infty$ the singlet $\tau_1^{s}(\xi),$ diverges with $\xi$. 
Of course for this latter state, since ${\bf{L}}_{\xi=\infty}(|\psi^{s}\rangle\langle\psi^s|)=0,$
all the $\tau_n$'s  diverge.
\begin{figure}
\begin{center}
\input{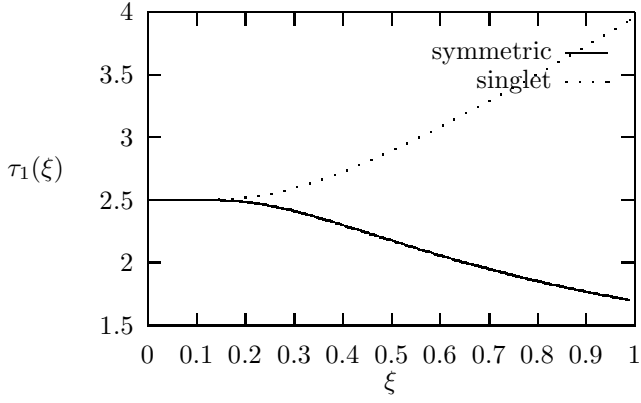}
\end{center}
\caption{
First order time-scale for the symmetric state
$|\psi^{sym}\rangle= (S^+)^2\,|0\rangle,$
 and a singlet state, $N=4$
$\Gamma(i,\,j)=0.1\,e^{-|i-j|/\xi}$.
}
\protect\label{Fig1}
\end{figure}
In the general case when the matrices ${\bf{\Gamma}}^{(\pm)}$ are not block constant
one has to resort to numerical calculations.
We have solved equation (\ref{ME})
by direct numerical integration  in the qubit case
with $H_{\cal R}=\epsilon\,S^z.$ 
Rather than using the form (\ref{explicit}) for the ME parameters, we
have choosen a phenomenological parametrization such as 
$\Gamma_{ij}^{(\pm)}=\Gamma_0^{(\pm)}\,e^{-|i-j|/\xi_c}$ and neglected
the self-Hamiltonian renormalization.
 In figure (\ref{Fig1}) is reported the behaviour of
 $\tau_1(\xi_c)$  for a $N=4$ singlet  and  the highest-weight $su(2)$-vector
belonging the $S=2$ multiplet. We see that for a wide range of $\xi_c$
the decoherence time of the singlet state   is  much 
 larger than that  of the symmetric state.
\begin{figure}
\begin{center}
\input{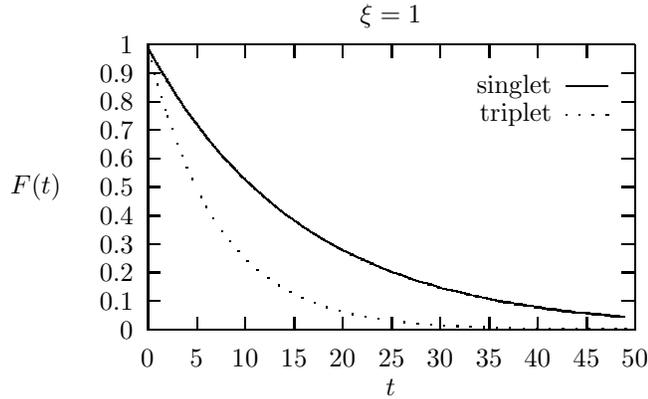}
\end{center}
\caption{
Fidelity as a function of the time for the $S^z=0$
singlet and triplet state
($N=2, \, \Gamma_0=0.1\,e^{-|i-j|/\xi}$)
}
\protect\label{Fig2}
\end{figure}
\begin{figure}
\begin{center}
\input{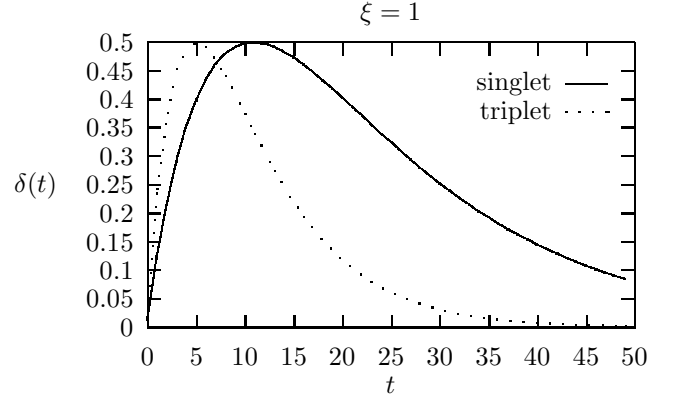}
\end{center}
\caption{
Linear entropy as a function of the time for the $S^z=0$
singlet and triplet state
($N=2, \, \Gamma_0=0.1\,e^{-|i-j|/\xi}$)
}
\protect\label{Fig3}
\end{figure}
\begin{figure}
\begin{center}
\input{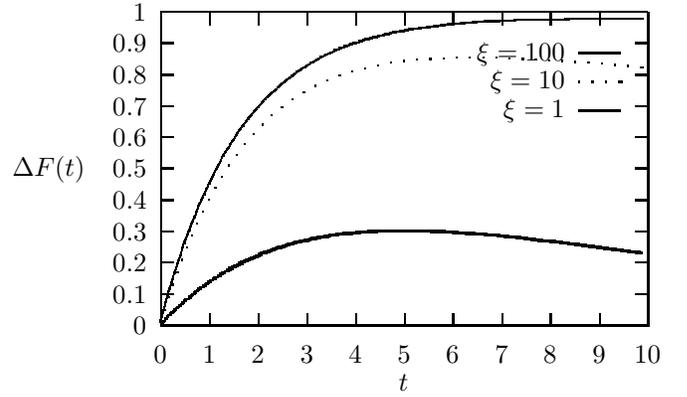}
\end{center}
\caption{
 Fidelity difference,  between a $N=4$ singlet and the state $(S^+)^2\,|0\rangle,$ 
 for different bath coherence lengths $\xi$
 ($\Gamma_0=0.1\,e^{-|i-j|/\xi}.$)
}
\protect\label{Fig4}
\end{figure}
\begin{figure}
\begin{center}
\input{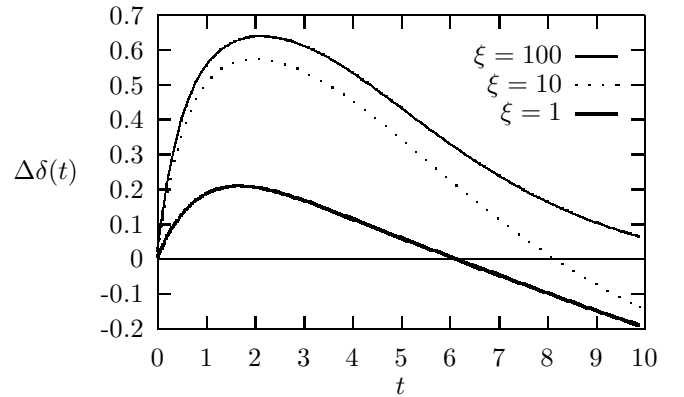}
\end{center}
\caption{
 Linear entropy difference,  between the state $(S^+)^2\,|0\rangle,$ and a  $N=4$ singlet,
 for different bath coherence lengths $\xi$
 ($\Gamma_0=0.1\,e^{-|i-j|/\xi}.$)
}
\protect\label{Fig5}
\end{figure}
In figures (\ref{Fig2}), (\ref{Fig3})  is compared the behaviour of the fidelity and   linear entropy
of the $N=2$ singlet and ($S^z=0$) triplet states at finite $\xi_c.$
Figures (\ref{Fig4}) and (\ref{Fig5}) show, as a function of time,
    the difference of fidelity and linear entropy, between   one
  of the $N=4$ singlets and the symmetric state $(S^+)^2\,|0\rangle \in{\cal H}_1(2),$
  for various bath coherence length.
These simple calculations  strongly suggest that the {\sl  noiseless
 enconding at infinite coherence length remains,  for sufficiently large  $\xi_c,$
more robust than all other states}
\subsection{ Gauge transformation}
We end this section by showing that  for a class of non-trivial qubit couplings, connected
to the  limits ii) and iii) via a  local gauge transformation, it  is possible
to build  subspaces annihilated by the dissipative component of the Liouvillian.
This transformation is a generalization of the one considered in \cite{DICKE}.
Here we give the proof for the replica-symmetric case, the cluster case 
being  a straightforward generalization.
Let us suppose that $\Gamma^{(\pm)}_{ij}=\Gamma^{(\pm)} \,e^{i\,(\phi(i)-\phi(j))},$ 
where $\Gamma^{(\pm)},\in{\bf{R}},\,\phi\colon {\bf{N}}_N\rightarrow {\bf{R}}.$
This kind of situation is not completely fictitious:  for  $g_{ki}\sim e^{i\,k\,{\bf {r}}_i}$
when 
there is just one bath mode $k$ degenerate with the qubit energy $\epsilon,$ 
from the first of equations (\ref{explicit}) follows that $\Gamma_{ij}^{(\pm)}\sim  e^{i\,k\,({\bf {r}}_i-{\bf {r}}_j)}.$
Introducing the operators $L^\sigma_{\phi} =\sum_{j=1}^N e^{i\,\phi(j)} A^\sigma_j$
The dissipative  Liouvillian has the canonical form (\ref{lindblad})
with $\{\lambda_\mu^\sigma\}=\{\Gamma^{(\sigma)}\}$ and Lindblad operators
given by the $L^\sigma_{\phi}$'s.
The operators $\{L^\sigma_\phi\}_\sigma,$ 
spanning a Lie Algebra ${\cal A}_\phi$ isomorphic to $\cal A$ generated by $\{L^\sigma\}_\sigma,$
are obtained from the latter 
by means of the (local) $U(1)$ gauge transformation 
\begin{eqnarray}
T_\phi &\colon&  \mbox{End} ({\cal H}_{\cal R})\rightarrow
\mbox{End} ( {\cal H}_{\cal R}) \colon X\rightarrow U_\phi\,X\,U_\phi^\dagger,\nonumber \\
U_\phi &=& \exp\{i\,\epsilon^{-1}\,\sum_{j=i}^N \phi(j)\,H^C_j\}\in\otimes_{j=1}^N U(1)_j,
\end{eqnarray}
where we recall that $H^C_i$ is the single cell Hamiltonian
fulfilling relation (\ref{algebra}) with the $A_i$'s.
The unitary operator $U\in\mbox{End} ({\cal H}_{\cal R})$ maps
the singlet sector ${\cal C}$ of ${\cal A}$ onto the one of $\tilde{\cal A}_\phi.$
Therefore  $\rho\in{\cal C}\Rightarrow \tilde{\bf{L}}(T_\phi \rho)=0.$
The new code $U_\phi({\cal C}),$ is noiseless depending  on the transformation properties of 
$H_{\cal R}^\prime $ under $T_\phi.$
If $H_{\cal R}= T_\phi(H_{\cal R}),$ (local gauge invariance)
it follows that  $U_\phi({\cal C})$ is noiseless under ${\bf{L}}_\phi$  {\sl iff}
${\cal C}$ is noiseless under ${\bf{L}}_{\phi=0}$ (replica independent case).
Let us consider, for example, the qubit case with $N=2$ and
 $\phi(j)=\phi\,j\,(\phi\in{\bf{R}})$
and $H_{\cal R}^\prime =\epsilon\,S^z.$
The noiseless state is now
the singlet $|\psi_s\rangle =2^{-1/2}(|\uparrow\downarrow\rangle-|\downarrow\uparrow\rangle).$ 
It is mapped by $ T_\phi$ 
onto $U_\phi|\psi_s\rangle=2^{-1/2}(e^{i\,\phi/2}|\uparrow\downarrow\rangle -e^{-i\,\phi/2}\,|\downarrow\uparrow\rangle),$
in particular for $\phi=\pi,$ one has $U_{\pi}|\psi_s\rangle=|\psi_t\rangle,$ that is
the triplet state becomes the noiseless one.
For $\phi\in(0,\,\pi)$ one has a smooth interpolation from $|\psi_s\rangle$ to $|\psi_t\rangle.$ 
It should be emphasized  that even if $ T_\phi(H_{\cal R})=H_{\cal R},$ 
generally one has that the 
many-qubit correction  $\delta H_{\cal R}$ is not invariant.
Nevertheless the  Hamiltonian part of ${\bf {L}},$ {\sl does not affect the
first order decoherence rate:  $U_\phi({\cal C})$  is sub-decoherent.}
\section{Conclusions}
In this paper we have studied a model of quantum register $\cal R$  with $N$ cells made of  
  replicas 
of a $d$-dimensional quantum system.
The register $\cal R$ is coupled with the environment, modelled by a thermal bath of harmonic oscillators,
through  single-cell operators $A_i$.
The latter are step operators over the spectrum of the cell Hamiltonian. 
The reduced dynamics of $\cal R$ is studied by a Master Equation  (ME) obtained in the Born-Markov 
approximation.
The ME provides a very natural and powerful tool
to discuss, in a unified way, the various aspects of  decoherence and dissipation
phenomena induced in $\cal R$ by the bath.
The effect of the environment splits into  two contributions:
a renormalization of the register self-Hamiltonian, that makes the cells effectively interacting,
and an irreversible component describing the decay processes.
The latter can be cast in canonical   Lindblad form by diagonalizing the $N\times N$ matrices
${\bf{\Gamma}}^{(\pm)}$; which  contain all  information about the effective spatial structure
of $\cal R$ in the given environment state.
Three situations which appear to be relevant for
 quantum encoding  have been discussed: i) all the cells are coupled
with the environment in the same way, ii) different cells feel
different  environments, iii) the register can be decomposed
in  uncorrelated clusters, such that the cells within each cluster satisfy i).  
In each of these cases one can show the existence of  subspaces $\cal C$
such that an initial pure preparation $|\psi\rangle\in{\cal C}$ 
has vanishing 
linear entropy production rate.
The states in $\cal C$ therefore -- on a short time-scale --
maintain  quantum coherence:  $\cal C$ can be thought of
as a subdecoherent  code.
The latter is obtained as simultaneous eigenspace  $\cal C$ of the Lindblad operators $L_\mu,$ 
given by linear
functions of the $A$'s associated with  the register cells.
Depending on the structure of the Lie algebra $\cal L$ generated
by the $L_\mu$'s one has to face  rather different situations.
For $A$ hermitian  $\cal L$ is abelian, 
the Hilbert space splits in a direct sum of  the simultaneous eigenspaces 
$\cal C:$ 
the ME is exactly solvable. Analytical expressions for  decoherences rate
can be found in the qubit case. 
In the non-hermitean case $\cal L$ is non abelian,
the Hilbert space splits according to the $\cal L$-irreps,
$\cal C$ is the common null space of the $L_\mu$'s (singlet sector of $\cal L$).
The latter exists, according to reference \cite{ZARA}, if the size of the clusters
satisfying i) is large enough.
For the qubit case the minimun cluster size required to encode one logical qubit is $N=4:$
a register made of $M$ clusters of  four qubit each supports a   $2^M$-dimensional subdecoherent space.
If $\cal C$ is left invariant by the renormalized self-Hamiltonian 
$H_{\cal R}^\prime$ of $\cal R$
the time-evolution of the sub-decoherent states is unitary: the code is noiseless.
In this case the relevant algebra is ${\cal L}^\prime$
generated by the Lindblad operators {\sl plus} $H_{\cal R}^\prime.$
Furthermore we have shown that there exist cases with non-trivial cell dependence
that can be mapped onto ii)-iii) by a suitable local gauge transformation.
The degree of stability of the resulting codes  depends on the covariance
properties of the renormalized self-Hamiltonian.
When the ${\bf{\Gamma}}^{(\pm)}$'s are not block-diagonal one has to resort to numerical calculations.
We have integrated the dissipative ME  of a qubit register. The results show that
for a wide range of bath coherence lengths $\xi_c$ the singlet states (noiseless at $\xi_c=
\infty$) are more robust, namely their entropy increases more slowly on the time scale
of decoherence.
\\
The problems related to the pratical realizations of the registers
satisfying the constraints for the suggested encodings, the preparation
as well as the gate manipulations of the codewords necessary in the quantum computation 
applications, are of course open issues that 
deserve further investigations.
\begin{acknowledgments}
The author thanks M. Rasetti 
for stimulating discussion and a careful reading of the manuscript;
 G.M.  Palma and C. Macchiavello for comments and suggestions.
Thanks are due to C.Calandra and G. Santoro for providing access to CICAIA
of Modena University and to Elsag-Bailey for financial support. 
\end{acknowledgments}

\end{multicols}

\begin{references}
\bibitem{NOTE} This result also follows simply by observing
in view of  Eq. (\ref{algebra}) and the 
finite-dimensionality of ${\cal H}_{\cal R}$,
 the $L_\mu^\sigma$'s must be nilpotent.
\bibitem{SYM}
More technically  it suffices that the $L_\mu^\sigma$'s and $H_{\cal R}$ belong 
to the $N$-fold tensor representation
of the {\sl universal enveloping algebra}  ${\cal U}(su(2)).$
\end{references}
\end{document}